# An Extended Classification and Comparison of NoSQL Big Data Models

**Sugam Sharma, PhD**

Center for Survey Statistics and Methodology, Iowa State University, Ames, Iowa, USA

Email: sugamsha@iastate.edu

**Abstract.** In last few years, the volume of the data has grown manyfold (beyond petabytes). The data storages have been inundated by various disparate potential data outlets, leading by social media such as Facebook, Twitter, etc. The existing data models are largely unable to illuminate the full potential of Big Data; the information that may serve as the key solution to several complex problems is left unexplored. The existing computation capacity falls short for the increasingly expanded storage capacity. The fast-paced volume expansion of the unorganized data entails a complete paradigm shift in new age data computation and witnesses the evolution of new capable data engineering techniques such as capture, curation, visualization, analyses, etc. In this paper, we provide the first level classification for modern Big Data models. Some of the leading NoSQL (largely being translated as "not only SQL") representatives of each classification that claim to best process the Big Data in reliable and efficient way are also discussed. Also, the classification is further strengthened by the intra-class and inter-class comparisons and discussions of the undertaken NoSQL Big Data models.

## 1 Introduction

The large, complex, heterogeneous, structured or unstructured data, under the ambit of Big Data has gained considerable attention and momentum in last two years. The rampant usage of social media such Twitter, Facebook, etc. is majorly responsible for such a severe data proliferation. The total size of the data, contributed by the social media itself is approximately 90% of the total data available and such data is sparsely located in disparate formats. In its informal definition, Big Data may consist of large, diverse, complex, and distributed, structured or unstructured data. The data can be gathered from a number of potential resources (Villars et al., 2011; IBM, 2012) such as instruments, sensors, internet transactions, email, video, click streams, etc. Big Data (White, 2012; Robert, 2012) is evolving as a predominant research area that requires robust data engineering capabilities in place to make effective advances in data science. The volume expansion of Big Data with such a fast pace imposes big research challenges. The data size experiences an increase of order of magnitude and in year 2012 alone, it grows from a few dozen terabytes to many petabytes in a single data set (Watters, 2010). The volume expansion of the data in its breadth and depth is much faster than the management techniques, required to handle such mammoth data; and, this results in the paucity of robust and appropriate data engineering systems. Most of the data processing systems still rely on their legacies of 1970s, and 1980s, despite the significant remodeling endeavors from time to time (Carino and Sterling, 1998). The timely reviews on computing solutions for handling the mammoth high-dimensional data sets, especially in bioinformatics (Schadt et al., 2010) offer some early insights of impending Big Data era. The reviews suggest that the existing technologies -- cloud and heterogeneous computing -- for years, could be the potential solutions, but question about their computing capacities also. As, their usages not only increase additional complexity, but also they could not sustain the stress of large amount of data, travels between the computing nodes and often break down.

In this Big Data era, the growing data proliferation challenges, the decreasing data engineering belief in the traditional data management and querying techniques, the poky emergence of robust and efficient data models, and sturdy motivation for data-driven applications are some of the critical concerns that ungently require prompt and collaborative actions from various scientific communities. Today, in scientific and industrial institutions, the term "data" is gradually ceasing its sovereignty to a newly popular term "Big Data", which is being ubiquitous in scientific community, academia and industry.

The inception of Big Data has revolutionized the traditional systems in academia or industry that engage the large presence of the data in their operations. In this data dominance era, data-driven approaches are evolving and widely being adopted in communities and adjoining the empirical, theoretical, and computational (Bell et al., 2009), a new paradigm of science is emerging and is being termed as "Data Science". Unlike its predecessors, in Data Science era, if the advantages are many, then the challenges are many too and one of the biggest challenges is the hardware infrastructure, housing very high end computation sources that are capable of sucking in Big Data successfully in making new scientific discoveries. Over, the last few years, the number of internet users, especially the social media, has grown in many-folds that result in massive and complex data (more data) production at a very fast-pace. This requires more efficient query (more queries) development that retrieves most accurate information (better results) with best optimal latency. The challenges demand the expansion of the existing infrastructure to the large, efficient and scalable infrastructure. Figure 1 demonstrates the comparison of growth between the data size and the processing capacity available annually; the calculations behind this juxtaposition are based upon the facts provided by Hilbert and López (2011) in their published work in Science. It can be observed how state-of-the-art computation capacity is lagging behind in front of the speedily leading data size, which is available to discover valuable knowledge patterns.

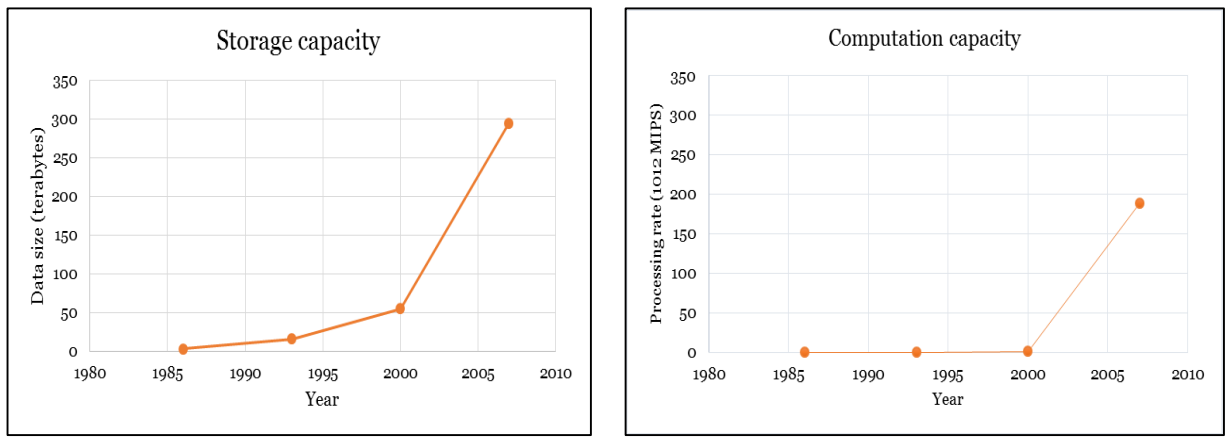

**Figure 1** Storage capacity vs. Computation Capacity: Data size exceeds the computation capacity (Source: Bell et al., 2009)

Rest of the paper is structured as follows. Section 2 provides a brief overview of Big Data. Section 3 is the classification and comparison of Big Data models. This section also reviews and discusses in detail some of the leading NoSQL Big Data models that address the Big Data challenges. The data models discussed under each classification are also compared on some salient features. Section 4 provides the detailed discussions on the considered classifications of the Big Data models. Finally, the paper is concluded in section 5.

## 2  Brief Summary of Big Data

The age of data science is gradually but rapidly evolving in most scientific communities. Various leading data outlets such as Facebook, Twitter, Wikipedia, Google, etc. are responsible for the massive deposit of the digital footprints and sizeable traces of the data. In last few years, due to the internet sites, the sources of data collection have been increased. The data is inundated from many potential sources such as sensors - gathering climatic information, power meter readings, social media sites, videos and digital pictures, traffic data etc. These sources are responsible for rapid growth of voluminous data - about 2.5 quintillion bytes every day (IBM, 2012). The rapid growth of heterogeneous data on such a large scale has infused some traces of doubts on the computing capacity of the existing data models. Because of the increasing of the data with such a fast velocity, the size of data is a centrally moving issue today. In 2012, in a single dataset, the size of the data likely to increase from few dozen terabytes to many petabytes and about 90% of the whole data in world today is produced in last two years only (IBM, 2012). Not only from storage capacity, but also from processing aspects, the existing database management systems fall short to handle them. Numerous scientific communities are interested to analyze such data to procure the useful findings. Though, the scientists are optimistic about the Big Data, but express equally low confidence in the data access capabilities of the existing techniques. They strongly advocate the development of new technologies to cater the current data processing need. In this era of data science, in its entirety, the term "large, complex, heterogonous, structure or unstructured data" has been redefined today and popularly is being called as "Big Data". Today, the importance and concerns of the data have been surpassed the challenges of the associated operations; in view of this, we try to redefine the operational-oriented definition of "Big Science" (Furner, 2003) into its data-centric equivalent and rephrase as –"The science that deals with Big Data is Big Science." The state of the art research in Big Data is still far behind from it maturity, has enormous scope of demystification and seeks continuously active research engagement from disparate scientific communities in order to derive new findings. The White House also has recently announced a national "Big Data" initiative, aiming to develop new core techniques and technologies to extract and use the knowledge from collections of large data sets to enhance the scientific research and innovation of managing, analyzing, visualizing, and extracting useful information from large, diverse, distributed and heterogeneous datasets (Weiss and Zgorski, 2012). This will increase the understanding of human and social processes and interactions and promote economic growth and improved health and quality of life. This section highlights also some of the important concerns about NoSQL data, which are briefly described here.

## 3 NoSQL Data Models

In this section, we briefly review some of the data models that are actively being discussed in NoSQL (of petabyte size) arena and are considered as the reliable, efficient and leading data models that defecate the NoSQL data stress. An ideal NoSQL model is expected to have the following attributes (Han, Song, and Song, 2011; Fan, 2010) -- *high concurrency, low latency, efficient storage, high scalability, high availability, reduced management and operational costs*. There are numerous modi operandi that help categorizing the NoSQL data models. Though, each category may have subcategories. In this section, the models are grouped together based on their classification and are discussed below. Under each classification, the mentioned date models are briefly discussed also and their current popularity is given special emphasis and based upon the popularity, the discussed models are ranked. The calculation of the popularity value is based upon the mechanism used by solidIT (2014), which counts the usage of the name of the model on various web spaces to derive the conclusion about the popularity ranking such as frequency of searches in Google Trends, count of the number, the model name is used as word in the tweets on the Twitter, etc.

### 3.1 Document-Oriented Store

The idea behind the concept of document store (Cattell, 2011) classification is a "document". Though, the underneath implementation may vary for individual data model, but in general, for this category, the data, formatted or encoded in XML, YAML, JSON, or BSON, etc., is encapsulated in documents and organization of the documents is implementation dependent. In the storage, each document is assigned a unique key that serves as the unique identifier for that specific document and data models are equipped with APIs or some query processing system to access the documents. The important data models of this category include MongoDB, CouchDB, OrientDB, Couchbase, MarkLogic, RavenDB, Cloudant, GemFire, etc. and some of popular models of this classification are briefly discussed here.

### 3.1.1 MongoDB

MongoDB (term extracted from the word hu**mongo**us) is an open source, document-oriented NoSQL database that has lately attained some space in the data industry (Chodorow and Dirolf, 2010). It is considered as one of the most popular NoSQL databases, competing today and favors master-slave replication. The role of master is to perform reads and writes whereas the slave confines to copy the data received from master, to perform the read operation, and backup the data. The slaves do not participate in write operations but may select an alternate master in case of the current master failure. MongoDB uses binary format of JSON-like documents underneath and believes in dynamic schemas, unlike the traditional relational databases. The query system of MongoDB can return particular fields and query set compass search by fields, range queries, regular expression search, etc. and may include the user-defined complex JavaScript functions. As hinted already, MongoDB practice flexible schema and the document structure in a grouping, called Collection, may vary and common fields of various documents in a collection can have disparate types of the data.

**Example query 1: Express the query in MangoDB that retrieves all the documents in the collection where the value in [qty] field is greater than 200 or the [price] field value is less than 10.**

In MangoDB, the query document (in JSON format) can be expressed as follows.

```
db.inventory.find ({$or:[{qty:{$gt:200}},{price:{$lt:10}}]})
```

The MongoDB is equipped with the suitable drivers for most of the programming languages, which are used to develop the customized systems that use MongoDB as their backend player. There is an increasingly demand of using MongoDB as pure in-memory database; in such cases, the application dataset will always be small. Though, it is probably is easy for maintenance and can make a database developer happier; this can be a bottle neck for complex applications that require tremendous database management capabilities.

Some of the prominent users of MongoDB are MetLife, Craigslist, Forbes, The New York Times, Sourceforge, eBay, etc. For example, The New York Times has its own form-building application that allows photo submission. This application is complete backed up by MongoDB. Sourceforge, on the other hand, has shown more confidence in MongoDB and use to store back-end pages. eBay Inc. uses it for the search suggestion application as well as for its own internal Cloud Manager.

### 3.1.2 CouchDB

CouchDB (Cluster of Unreliable Commodity Hardware Data Base) is an open source, NoSQL, web inclined database that uses JSON to store the data and relies on JavaScript as its query language (Brown, 2011).

CouchDB is equipped with a built-in administration web interface, called Futon (Anderson, Slater, and Lehnardt, 2009). As stated already, that APIs are used to access the data in CouchDB, a few of the example queries are expressed here using the command-line utility curl. By default, it issues a GET requests only, but POST requests can be issued also, but it requires explicit mentioning about it.

**Example query 1: Create a database, called baseball.**

```
curl -X PUT http://127.0.0.1:5984/baseball
The CouchDB replies in JSON format as {"ok": true}.
```

**Example query 2: Retrieve the list of all databases.**

```
curl -X GET http://127.0.0.1:5984/_all_dbs
The CouchDB responds as ["baseball"]
```

**Example query 3: Create a document that consists of the price of apple at different supermarket stores.**

The document is created using Futon and document structure looks like this. Here, the _id and _rev fields are generated by CouchDB itself.

```
{
   "_id": "00a271787f89c0ef2e10e88a0c0001f5",
   "_rev": "1-2628a75ac8c3abfffc8f6e30c9958fd7",
   "item": "apple",
   "prices": {
       "Fresh Mart": 3.1,
       "Price Max": 3.5,
       "Apples Express": 3
   }}
```

The usability of CouchDB is growing and its prominent users include Ubuntu, The British Broadcasting Corporation (BBC), Credit Suisse, Meebo, etc. Ubuntu used it for the in-house synchronization service, but relinquished its use later in 2011. BBC exploits the computational capacity of CouchDB for the platforms used for dynamic content. Credit Suisse is using CouchDB for their internal commodity departments.

**Discussion.** Under the document store category, the two most popular data models are discussed and figure 2 depicts the popularity trends for MongoDB and CouchDB (solidIT, 2014). It can be noticed that MongoDB always enjoys bit higher popularity trends compared to CouchDB. The important features of the MongoDB and CouchDB are shown in table 1. The features of the data models are also compared. The importance of indexing (Bertino et al., 2012) in database has been proven very efficient since a long period of time, particularly when data size is huge. Indexing in Big Data is a challenge comparatively due to several reasons. Primarily, the volume and the velocity of the Big Data are larger to the order of magnitude comparatively; the generated index should be smaller (a fraction of original data) and equally faster as compared to the data as the search is conducted through billions or even trillions of data values in seconds. In MongoDB, indexes (Chodorow and Dirolf, 2010), which use B-tree data structure, are generated at the collection (equivalent to a table in relation databases) level. To support variety of data and queries, there are different types of index are available in MongoDB: 1) Default_id - this exists by default on _id field of collections and also serves as unique id too, 2) Single Field - this is a user defined index on a single field of a document, 3) Compound Index - this is user defined index on multiple fields, 4) Multikey Index - this is used to index a field that hold an array of value, 5) Geospatial Index - this is a very special index that MongoDB (unlike other models) provides to support spatial queries and are of two types: (i) 2d spatial index for planer geometry queries, (ii) 2sphere spatial index for spherical geometry queries, 6) Text Index - this is the index type that helps searching for the string or text content in a collection, 7) Hashed Index- this helps indexing the hash of the values of a field.

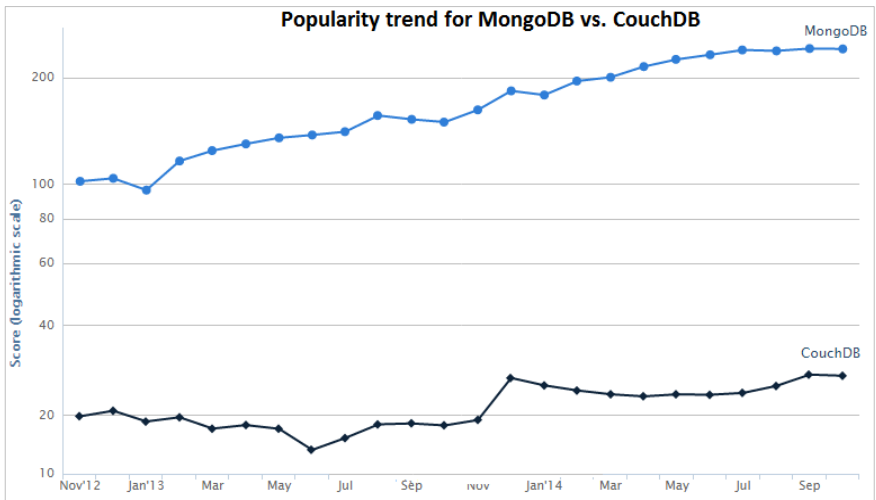

**Figure 2** Popularity trend for MongoDB vs. CouchDB (Source: solidIT, 2014)

**Table 1** Feature comparison of MongoDB vs. CouchDB

| Model / Features | MongoDB (Year: 2007) | CouchDB (Year: 2005) |
|---|---|---|
| **Relational nature** | No | No |
| **Developer** | 10gen | IBM (Damien Katz) |
| **Written in** | C++ | Erlang |
| **Query language** | MongoDB ad-hoc query language | JavaScript |
| **SQL nature** | No | No |
| **High availability** | Yes | Yes |
| **High scalability** | Yes | Yes |
| **Single point of failure** | No | No |
| **Open source** | Yes | Yes |
| **Versioning** | Yes (different database for audit trails, not native) | Yes (MVCC) |
| **Indexing** | Advance and wide variety (includes spatial indexing) | Basic associated map and reduce functions |
| **Data Processing nature** | Batch processing and event streaming | Batch processing |

CouchDB on the other hand does not infuse the power of indexing directly, unlike relational database models. In contrast, the usage of index in CouchDB is assisted by map and reduce functions of the predefined MapReduce framework (Ranger et al., 2007); any variations in the document structure of the data are well absorbed by these functions and encourage parallelism and independent computation in indexes too. The indexes created by map functions limit the window of the targeted data, the reduce queries are allowed to run to seek optimal efficiency. So, while writing the map function, building the index should also be a concern. In CouchDB, the amalgamation of a map and a reduce function is termed as a view. In sum, if we compare the indexing richness for MongoDB and CouchDB, former distantly leads and clearly is a preferred choice for better indexing.

Versioning (Shaughnessy, 2012), also termed as version control, revision control, is another important features that worth discussion. Versioning allows fetching latest or any particular version out from the tables in databases and also significantly eases the audit trailing of the data in the systems. MongoDB versions the document, through this is not its native feature but happens at application layer. The efficient way suggested is to have two different databases, where one is the application data store and other is audit trail. Though, CouchDB enjoys Multi Versioning Concurrency Control (MVCC) (Suryavanshi and Yadav, 2012; Cattell, 2011), but that does not warrant for the default document versioning support. It does implement a form of MVCC, especially to preclude the strict requirements of exclusive locking on data files while performing the write operation. Version conflicts are can be resolved at the application level by merging the conflicted data into one of the file, followed by the eradication of the rancid part.

### 3.2 Graph Data Model

The graph data models constitute and store the data along the nodes, edges, and properties of a graph structure (Angles and Gutierrez, 2008) and provide index-independent adjacency, where every element is connected to its adjacent elements through the direct pointers. This avoids any lookup for indexing. Some graph data models are Neo4J, OrientDB, InfiniteGraph, Allegro, Virtuoso, and Stardog. In this section, we also briefly discuss a few popular data models --*Neo4J, and OrientDB* -- of this category.

#### 3.2.1 Neo4j

Today, Neo4j is being considered as the world's leading graph data model and is a potential member of NoSQL family (Hoff, 2009). It is an open-source, robust in nature, disk-based graph data model that fully supports ACID transactions and implemented in Java. The graph nature of Neo4j attributes it with agility and speediness and in comparison to relational databases, for the similar set of operations; it is cited significantly faster and outperforms the former with greater than 1000x for several potentially important real-time scenarios (Eifrem, 2009). Similar to an ordinary property (simple key-value pairs) graph, Neo4j graph data model is consists of nodes and edges, where every node represents an entity and an edge between two nodes corresponds to the relationship between those attached entities. As location has become an important aspect of the data today, most of the applications have to deal with the highly associated data, forming a network (or graph); social networking sites are the obvious examples of such applications. Unlike relational database models; which require upfront schemas that restrict the absorption of the agile and ad-hoc data, Neo4j is a schema-less data model, which works on bottom-up approach that allows easy expansion of the database to imbibe ad-hoc and dynamic data. Similar to any other data model, Neo4j has its own declarative and expressive query language, called Cypher (Johnson, 2010) that has the pattern matching capabilities among the nodes and relationship during the data mining and data updating.

```
START: Starting points in the graph, obtained via index lookups or by element IDs
MATCH: Pattern matching, bound to the starting points in START
WHERE: Clause to filter the outcome
RETURN: Returning output
CREATE: Helps in creating nodes and relationships
DELETE: Helps in removing nodes, relationships and properties
SET: Setting values to the properties
FOREACH: Iterating through the list
```

```
     WITH: Query refactoring into multiple and distinct parts
```

**Example query 1.1: Start from all node in a graph.**
```
     START n=node (*);
```

**Example query 1.2: Start from one or more nodes in a graph, specified by ids.**
```
     START n=node ({ids})
```

**Example query 1.3: Multiple starting points in a graph.**
```
     START n=node ({id1}), m=node ({id2})
```

**Example query 1.4: Return all the nodes in the graph database.**
```
     MATCH (n)
     RETURN n
```

**Example query 2: Find a user called John, then traverse the graph looking for friends of John's friend (though not his direct friends) before returning both John and any friends-of-friends that are found.**

```
     MATCH (user {name: 'John'})-[:friend]
          ->()-[:friend]->(fnd-of-fnd)
     RETURN user, fnd-of-fnd
```

| User | fnd-of-fnd |
|---|---|
| Node[3]{name:"John"} | Node[1]{name:"Maria"} |
| Node[3]{name:"John"} | Node[2]{name:"Steve"} |
| 2 rows | |

**Example query 3: Traverse the graph to look for those users that have an outgoing friend relationship, returning only those followed users who have a name property starting with S.**

```
     MATCH (user)-[:friend]->(follower)
     WHERE user.name IN ['Joe', 'John', 'Sara', 'Maria', 'Steve'] AND follower.name =~ 'S.*'
     RETURN user, follower.name
```

| user | follower.name |
|---|---|
| Node[3]{name:"John"} | "Sara" |
| Node[4]{name:"Joe"} | "Steve" |
| 2 rows | |

### 3.2.2 OrientDB

OrientDB, developed in Java by Luca Garulli (2012) is an open source graph-document NoSQL data model. It largely extends the graph data model but conglomerates the features of both document and graph data models up to a certain extents. At the data level for the schema-less content, it is document-based in nature, whereas to traverse the relationship, it is graph-oriented and therefore, fully supports schema-less, schema-full or schema-mixed data. The database is completely distributed in nature that can be spanned across several servers. It supports the state of the art Multi-master replication distributed system. It is a full ACID compliance data model and also offers role based security profile to the users. One of the salient features of OrientDB is its fast indexing system for lookups and insertion and that is based on MVRB-Tree algorithm, originated from Red-Black Tree and B+ Tree.

OrientDB relies on SQL for basic operations and uses some graph operators extensions to avoid SQL joins in order to deal with relationships in data; particularly, Gremlin (2013), known as graph traversal language, is used as the query processing language and can loosely be termed as OrientDB's SQL.

**Example query 1: Report, per each food, the list of animals that eat that food.**

```
SELECT name, in.out.in.name FROM Food
Output: Meat, [Gaudì, Kelly]
```

**Example query 2: Extract the name of the animals that eat less than 1Kg of meat per day.**

```
SELECT name FROM Animal WHERE out.kgPerDay < 1 AND out[@class='Eat'].in.name = 'Meat'
Output: Gaudi
```

**Example query 3: Extract the name of the animals that eat at 10 AM.**

```
SELECT FROM Animal WHERE out[@class='Eat'].whenAsHours CONTAINS 10
Output:Kelly
```

**Example query 4: To extract all the outgoing vertices connected to 'Gaudì' Animal. The use of both SQL and GREMLIN is allowed.**

```
SELECT GREMLIN ("current.out") FROM Animal where name = 'Gaudì'
Output:Kelly
```

**Discussion.** We attempt to capture the popularity trends between Neo4j and OrientDB data models in figure 3 (solidIT, 2014). When popularity comparison is done between only these two data models, Neo4j is being discussed far more than OrientDB in literature. Though, Neo4j continuously enjoys its upscaled popularity since its inception; but, it can be clearly noticed in the trending graph that the wider gap of popularity between these two robust data models has been abated a bit lately. Table 2 is populated with some of the important, comparable, and self-descriptive features of Neo4j and OrientDB.

Neo4j does not have native versioning capacity rather it uses time-based versioning to version the data. OrientDB uses the semantic versioning (OSGi Alliance, 2010) in version assignment and tags the version in the X.Y.Z format, where X, Y, Z are integer, reflecting major, minor, and patch level values. As far as data processing is concerned, OrientDB primarily process the data in batch. Though, the data streaming is also feasible though streaming plugin.

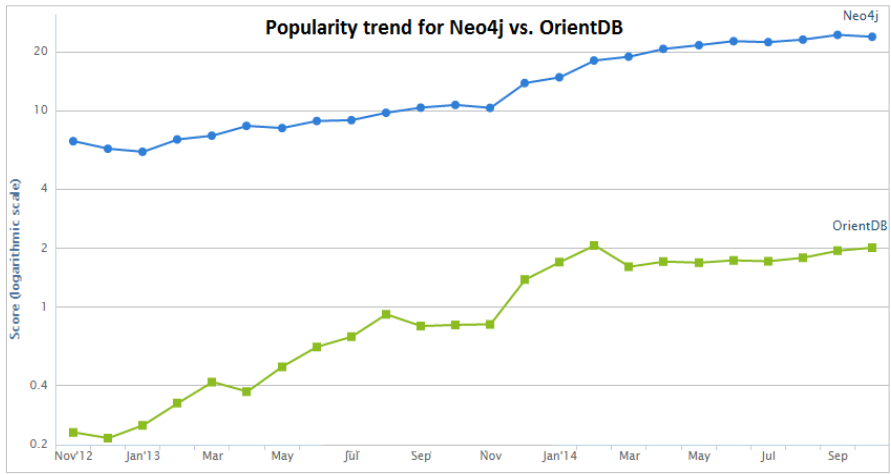

**Figure 3** Popularity trend for Neo4j vs. OrientDB

**Table 2** Feature comparison of Neo4j vs. OrientDB

| Model Features | Neo4j (Year: 2007) | OrientDB (Year: 2010) |
|---|---|---|
| **Relational nature** | No | No |
| **Developer** | Neo Technology, Inc. | Orient Technologies |
| **Written in** | Java | Java |
| **Query language** | Cypher | SQL (with some differences) |
| **SQL nature** | No | Yes |
| **High availability** | Yes | Yes |
| **High scalability** | Yes | Yes |
| **Single point of failure** | Yes | No |
| **Open source** | Yes | Yes |
| **Versioning** | Not native (Time-based versioning) | Yes (Semantic versioning) |
| **Indexing** | Legacy indexing, Lucene | SB-Tree, HashIndex, Lucene |
| **Data processing nature** | Batch processing | Batch processing primarily |

## 3.3 Key-value Store

At the fundamental level, a key-value store (Berezecki et al., 2011) can be built by an associative array -- map or dictionary. Such data model, stores the data as the collection of key-value pairs, where key serves as the primary key that cannot be duplicated in the same collection. The key-value data model is one of the most basic non-trivial data models that underpin the development of more complex data models. In this section, we discuss some of the popular key-value data models.

### 3.3.1 Riak

Riak is a key-value open-source NoSQL data model that has been developed by Basho Technologies (Collins et al., 2010). The model is considered as a fine implementation of the Amazon's Dynamo principles (DeCandia et al., 2007) and distributes the data through nodes by employing the consistent hashing in an ordinary key-value system into buckets, simply the namespaces.

**Example 1: Retrieve a specific key from a bucket**

- `GET /types/TYPE/buckets/BUCKET/keys/KEY;`

**Example 2: Perform a read operation on the key *rufus* in the bucket dogs, which bears the type animals.**

- `Location myKey = new Location (new Namespace ("animals", "dogs"), "rufus");`

### 3.3.2 Oracle NoSQL

In order to efficiently address the challenges of Big Data, the leading data vendor, Oracle has developed Oracle NoSQL database (Larry, 2012). It has been built by the Oracle Berkeley DB team (Abidi, 2011) and the Berkeley DB Java edition is the building block of Oracle NoSQL. Berkeley DB is a robust and scalable key-value store and used as the underlying storage for several popular data model such as Amazon Dynamo, GenieDB, MemcacheDB and Voldemort (Tiwari, 2011). There are several other database systems that discern the strength of Berkeley DB and have attained greater scalability, throughput, and reliability with little tuning efforts. It is an efficient and a resilient transaction model that significantly eases the development process of applications, involving Big Data. It is a distributed, scalable yet simple key-value pair data model that fully supports the ACID transactions and JSON format and integrated with Oracle Database and Hadoop. It offers scalable throughput with bounded latency.

**Example query 1: CRUD examples**

- Put a new key/value pair in the database, if key not already present.
  ```
  Key key = Key.createKey("Katana");
  String valString = "sword";
  store.putIfAbsent(key, Value.createValue(valString.getBytes()));
  ```
- Read the value back from the database.
  ```
  ValueVersion retValue = store.get(key);
  ```
- Update this item, only if the current version matches the version I read.
- In conjunction with the previous get, this implements a read-modify-write
  ```
  String newvalString = "Really nice sword";
  Value newval = Value.createValue(newvalString.getBytes());
  store.putIfVersion(key, newval, retValue.getVersion());
  ```
- Finally, (unconditionally) delete this key/value pair from the database.
  ```
  store.delete(key);
  ```

**Example query 2: Unordered Iteration Example**

```
▪ Create Iterator.
    Iterator<KeyValueVersion> iter = store.storeIterator(Direction.UNORDERED, 100);
▪ Now, iterate over the store.
    while (iter.hasNext()) {
        KeyValueVersion keyVV = iter.next();
        Value val = keyVV.getValue();
        Key key = keyVV.getKey();
        System.out.println(val.toString() + " " + key.toString() + "\n");
    }
```

**Example query 3: Wrapping a Sequence of Operations in a Transaction**

```
▪ Create a sequence of operations.
    OperationFactory of = store.getOperationFactory();
    List<Operation> opList = new ArrayList<Operation>();
▪ Create major and minor path components.
    List<String> majorComponents = new ArrayList<String>();
    List<String> minorLength = new ArrayList<String>();
    List<String> minorYear = new ArrayList<String>();
    majorComponents.add("Katana");
    minorLength.add("length");
    minorYear.add("year");
    Key key1 = Key.createKey(majorComponents, minorLength);
    Key key2 = Key.createKey(majorComponents, minorYear);
▪ Now put operations in an opList.
    String lenVal = "37";
    String yearVal = "1454";
    opList.add(of.createPut(key1, Value.createValue(lenVal.getBytes())));
    opList.add(of.createPut(key2, Value.createValue(yearVal.getBytes())));
▪ Now execute the operation list.
    store.execute(opList);
```

**Discussion.** We attempt to capture the popularity trends between Riak and Oracle NoSQL data models in figure 4 (solidIT, 2014). As the basic foundation of Oracle NoSQL is Berkeley DB, which is very widely used (as local Big Data model) even today, we try to mention it in the discussion, wherever possible. As it is already mentioned that Berkeley DB serves as the foundation for the newly developed Oracle NoSQL, Berkeley DB still maintains its popularity over Oracle NoSQL. Figure 4 also indicates that Riak is being discussed far more comparatively in literature. Though, the popularity trend for Berkeley DB is almost linear or rather slightly decreasing, a gradual demand for Riak and Oracle NoSQL can also be observed in figure 4. Table 3 contains some of the very important features of Riak and Oracle NoSQL; though, most of them are self-descriptive, we briefly discuss and compare a few of them.

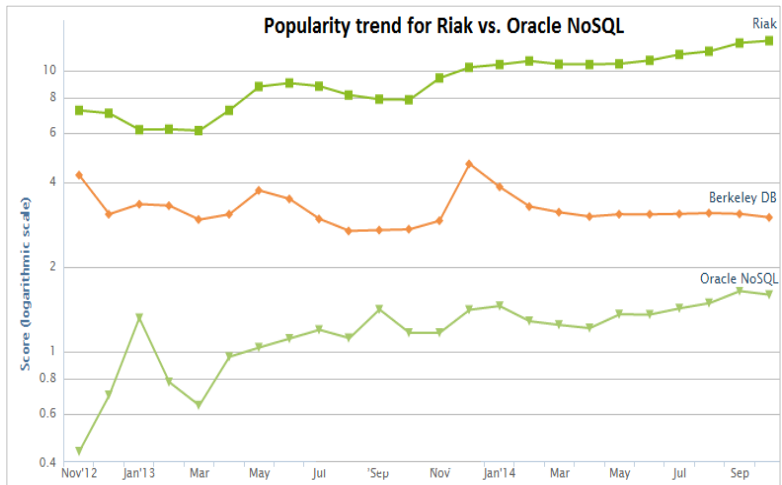

**Figure 4** Popularity trend for Riak vs. Oracle NoSQL and Berkeley DB

**Table 3** Feature comparison of Riak vs. Oracle NoSQL

| Model / Features | Riak (Year:2009) | Oracle NoSQL (Year:2011) |
|---|---|---|
| **Relational nature** | No | No |
| **Developer** | Basho Technologies | Oracle Corporation |
| **Written in** | Erlang | Java (also C library) |
| **Query language** | RESTful API | key access methods: put, get, delete JavaScript APIs |
| **SQL nature** | No | No (Mainly ) |
| **High availability** | Yes | Yes |
| **High scalability** | Yes | Yes |
| **Single point of failure** | No | No |
| **Open source** | Yes (also enterprise and cloud storage version) | Yes |
| **Versioning** | Yes (Vector clock) | Yes (Auto incremented integer number) |
| **Indexing** | Secondary indexing (2i) | secondary index |
| **Data processing nature** | Batch processing | Batch processing and Streaming |

## 3.4 Wide Column Store

Wide column stores (Cattell, 2011) at abstract level are key-value stores with two-dimensions. They are attributed with dynamic records that store the data and can house a huge number of dynamic columns; because of which, also termed as extensible record stores. In wide column stores, which enjoy schema-free characteristics, are column names are dynamic; so, the record keys and a record can house billions of columns. There are several data such models available today -- BigTable, Cassandra, Druid, HBase, Hypertable, etc. -- and the popular pool is discussed here.

### 3.4.1 BigTable

The paper by Chang et al. (2006) is one of the few introductory papers on BigTable. The paper outlines the simple data model of the BigTable alongwith the design and implementation. BigTable, primarily used by Google for various important projects, is envisioned as a distributed storage system, designed to manage the petabytes size of the data, distributed across several thousand commodity servers and allows further horizontal scaling.

In a write operation, a valid transaction is logged into the tablet log. The commit is the group commit to improve the throughput. On the commit moment of the transaction, the data is inserted into the memory table (memtable). During the write operation, the size of memtable continues to grow and once reaches the threshold, the current memtable freezes and a new memtable is generated. The former is converted into SSTable and finally written into GFS. In a read operation, the SSTables and memtable form an efficient merged view and the valid read operation is performed on it.

**Write operation:**

```
// Opening the table
Table *T = OpenOrDie("/bigtable/web/webtable");

// Writing a new anchor and deleting the old
RowMutation r1(T, "com.cnn.www");
r1.Set("anchor:www.c-span.org", "CNN");
r1.Delete("anchor:www.abc.com");
Operation op;
Apply(&op, &r1);
```

**Read operation:**

```
Scanner scanner(T);
ScanStream *stream;
stream = scanner.FetchColumnFamily("anchor");
stream->SetReturnAllVersions();
scanner.Lookup("com.cnn.www");
for (; !stream->Done(); stream->Next()) {
        printf("%s %s %lld %s\n",
                scanner.RowName(),
                stream->ColumnName(),
                stream->MicroTimestamp(),
                stream->Value());
}
```

The BigTable relies on Google's in-house developed, highly-available, loosely-coupled persistent distributed locking systems to synchronize accesses to the shared resources, called Chubby (Burrows, 2006). The service is developed to provide coarse-grained locking and offers a reliable distributed file system.

The BigTable Data Model is a high availability storage system that is highly scalable. It is a simple data model at the lowest end, can be imagined as multidimensional flat file. This technology is leveraged by more than 60 Google products through their APIs. At the fundamental hardware level, the distributed clusters are connected through the central switches and each cluster is constituted by various racks. A rack can consist of several commodity computing machines that communicate to each other though the rack switches. In BigTable data model the intra-rack communication i.e. the communication within a rack through the rack switches is very efficient.

In BigTable data model, there is no support for ACID (Atomicity, Consistency, Isolation, Durability) transactions across row keys. As, the BigTable data model is not relational in nature, it does not support join" operations and there is no SQL type language support also. This may pose challenges to those users, who largely rely on SQL like language for the data manipulation. In BigTable data model, at the fundamental hardware level, the inter-rack communication is a less efficient than in the intra-rack communication *i.e.* the communication within a rack through the rack switches. Though, the bottleneck is overcome through some software systems. The chubby lock service, used for shared resources synchronization has limitations on the scalability of the distributed file system.

The BigTables are used primarily by Google for its various applications such as Gmail (Copper, 2009), YouTube (Cordes, 2007), Orkut (Whitchcock, 2005), Google Code Hosting, Blogger.com (Google Inc., 2003), Google Earth (Google Inc., 2001), My Search History, Google Book Search (Google Inc., 2004), Google Maps (Whitchcock, 2005), Google Reader (Wetherell, 2005), web Indexing (Chang et al., 2006), and Map Reduce (Chang et al., 2006: p. 3).

### 3.4.2 Cassandra

Cassandra, originally devised by Lakshman and Malik (2011) is a distributed, open source data model that handles massive amount of data of petabytes magnitude, distributed across several commodity servers. The robust support, high data availability and no single point of failure by Cassandra is guaranteed by the fact that the data clusters span multiple datacenters. The Cassandra data model believes in parallel data operation that results in high throughput and low latency and adopts data replication strategy to ensure high availability for writing too. Unlike relational data models, in Cassandra the data is duplicated on multiple peer nodes that helps ensuring reliability and fault tolerance.

```
Example query 1:
      CREATE TABLE playlists(
      Id uuid,
      Song_order int,
      Song_id uuid,
      title text,
      album text,
      artist text,
      PRIMARY KEY (id, song_order));

Example query 2:
      INSERT INTO playlist (id, song_order, song_id, title, artist, album)
        VALUES (62c36092-82a1-3aoo-93d1-46196ee77204, 4,
                    7db1a490-5878-11e2-bcfd-o8oo2ooc9a66,
                    'ojo Rojo', 'Fu Manchu', 'No One Rides for Free');
Example query 3:
      SELECT * FROM playlists;
```

The durable write, set to true is advisable to avoid risk of losing data. Through customizable, but the right amount of memory is dynamically allocated by Cassandra system and a configurable threshold is set. During the writing operation when the contents exceed the limit of memtable, the data including indexes, is queued, ready to be flushed to SSTables (Sorted String Table) on disk through sequential I/O. The Cassandra model comes equipped with a SQL-like language called Cassandra Query Language (CQL) to query data. Example CQL queries are shown in here in this section (query 1, 2, and 3) and are self-descriptive.

Literature suggests that Cassandra does not have a nice PHP driver that can be used to program a system. There are some drivers that do not have proper documentation, on how to create a PHP extension for window application. There is an increasingly growing list of the prominent users that rely on the computational capabilities of Cassandra for their data management partially or fully; Facebook, IBM, Netflix, Rackspace, Twitter, etc. are some of them.

### 3.4.3 HBase

HBase, written in Java, is a non-relational, column-oriented, open source, and distributed database management system (Dimiduk and Khurana, 2012) that uses Hadoop Distributed File System (HDFS) underneath. It is designed to deal with very large sparse datasets and is a member of NoSQL family and follows the master-slave concept as well. Though, HBase is not a relational database and does not support SQL, but it is a column-oriented data model that consists of the sets of tables, consisting of rows and columns. Similar to relational database, in each table, there is a primary key that is used to access the data. HBase has the ability to group many columns (attributes) together into what are called column families and the elements of a column family are all stored together. Any new column can be added to a column family any time.

**HDFS.** The Hadoop Distributed File System (HDFS), written in Java programming language, is highly fault-tolerant file system primarily aimed to deploy and run on low-cost commodity hardware (Borthakur, 2007). It is very much appropriate for complex applications that deal with large datasets (up to terabytes) and streaming access with high throughput and scale to hundreds of nodes in a single cluster. HDFS believes on the principle that in case of extremely dataset, moving computation is cheaper than moving data. Moving the requested computation, which is relatively much smaller than the data it requires to operate on, execution near the data minimizes the network congestion up to a great extent and thus enhances the throughput of the system in its entirety.

HDFS is used as a bare base file system by HBase Data model (Dimiduk and Khurana, 2012) alone does not observe much usage, but in a complex package, known as Hadoop, with other component, it potential is enormous. Yahoo! Inc., Facebook, IBM, TIBCO, MetaScale, MapR Technologies Inc. are some of the vendors that either use or support it.

**MapReduce**. MapReduce is a programming model that has recently attained momentum in Big Data community (Ranger et al., 2007). It exploits parallel and distributed algorithms underneath on a cluster to efficiently process the humongous data and uses master-slave architecture underneath. As the name sounds, it is a tight conglomeration of two functions, called Map and Reduce and the model runs the various smaller tasks in parallel on multiple distributed servers. As a complete framework, the MapReduce model manages the data transfers and communication among different components of the system and takes care of the redundancy and fault tolerance challenges. The libraries of MapReduce are open-source and available in multiple languages; today, the most popular implementation Apache Hadoop (Borthakur, 2007).

```
Example query 1: Write the Map and Reduce functions that count the appearance of every word in
    a set of documents.

      function map(String name, String document):
        // name: name of the document
        // document: contents in the document
        for each word wo in document:
          outputFunc (wo, 1)

      function reduce(String word, Iterator wordCounts):
        // word: a word
        // wordCounts: a list of aggregated word counts
        TotalCount = 0
        for each wc in wordCounts:
          TotalCount += ParseInt(wc)
        outputFunc (word, TotalCount)
```

It can be noticed that both the functions have key-value pairs as the input parameters. The role of the key in Map function is to identify the data that is to be processed by the associated map instance. There may be several instances of the Map function poured across multiple machines. The Map function returns the output in the form of the key-value pairs which are used the Reduce function that reduces this large set of key-value pairs to desirably a single output, possessing the same input key, as a final result of the original complex problem.

MapReduce is extremely useful for a large pool of computing applications such as Google's index of the World Wide Web, distributed sorting, web access log stats, pattern-based searching, etc.

**Discussion.** Figure 5 shows the popularity trends for Cassandra, HBase, and BigTable (solidIT, 2014). It can be noticed that Cassandra has a bit higher popularity trends almost linearly compared to HBase. If the comparison is done among these three data models discussed in wide column category, though, BigTable is distantly far behind in popularity, but HBase is enjoys its popularity much closely with Cassandra, though, still stays at the lower end. Table 4 contains some of the important features of the Cassandra, HBase, and BigTable; the features are intuitive and self-descriptive.

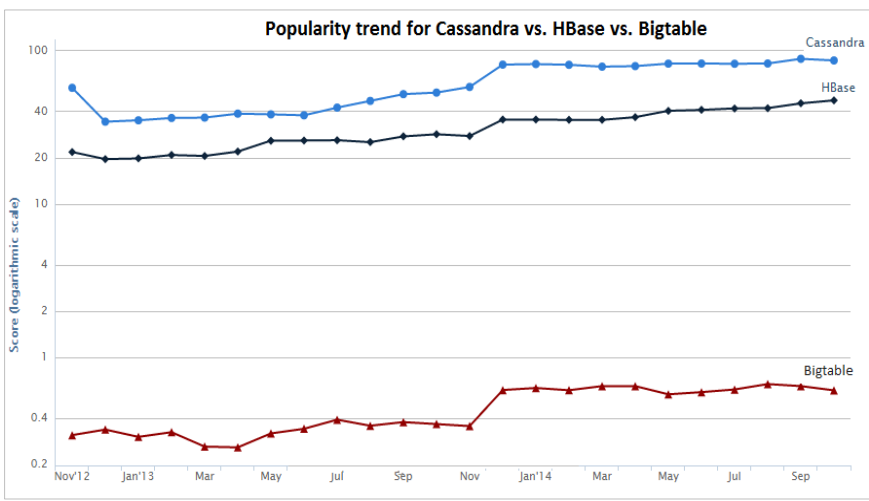

**Figure 5** Popularity trend for Cassandra vs. HBase vs. BigTable (Source: solidIT, 2014)

Unlike other traditional Big Data models, in HBase, MapReduce is a software programming model that is known for its efficient parallel and distributed computing capabilities. It becomes extraordinarily useful when it is convoluted with hardware of the data models on the top of them to serve as demons. As of late, the interweaving proximity of MapReduce programming model with Apache Hadoop framework has drawn significant attention and popularity in the recent buzz of Big Data.

Indexes support the efficient execution for more often used queries, especially for the read operations (Bertino et al., 2012). An index may include a column or more and sometimes the size of an index may grow larger than the table itself, it is being created for, but eventually provides the rapid lookup and fast access of the data and this compensates for the overhead of having indexes. As Big Data models entertain multi-variable queries, the indexing is required to be in place for all the in-operation variables, promising faster access for individual variable search results that are further combined for collective response as a whole. The model should have the ability to produce smaller indexes, when there is a possibility of granularity reduction. The indexes should able to be easily portioned into sections whenever the opportunities for parallel processing arise. As the rate of data generation under the Big Data ambit is much faster, for in-situ processing, the index should be built at the same rate. In Cassandra, similar to the relational database where the first index is the primary key that ensures the uniqueness of records, the primary index, applied to a column family, serves as the index for its row keys and is maintained by each node for its associated data (Lakshman, and Malik, 2011). Cassandra also provides secondary indexes on column values which can act as additional filter to the resultset. HBase is not equipped with robust indexing capability (Dimiduk and Khurana, 2012). Secondary indexes are not built-in by default, but, can be generated in other table that may require periodic updated and a MapReduce job performs the task; another possibility is to write to the index table while writing to the data table. Through, each approach has both merits and demerits, but, all in all HBase does not aggressively support indexing. Like HBase, BigTable data model also is not a good contender of indexing (Jin et al., 2011). As, it is already stated that technically, BigTable is a map, which is sparse, distributed, persistent, multidimensional and sorted. Though, the map is indexed by a row key, column key, and a timestamp.

As far as versioning (Shaughnessy, 2012) issue is concerned, in HBase, the version management of the data is assisted by the timestamps.

**Table 4** Feature comparison of Cassandra vs. HBase vs. Bigtable

| Model Features | Cassandra (Year: 2007) | HBase (Year:2007) | BigTable (Year:2004) |
|---|---|---|---|
| **Relational nature** | Yes | No | No |
| **Developer** | Facebook Inc. (Avinash Lakshman, Prashant Malik) | Microsoft (Powerset) | Google Inc. |
| **Written in** | Java | Java | C, C++ |
| **Query language** | CQL | 1.Pig latin 2. HQL | APIs in C++ |
| **SQL nature** | Yes | 1.No 2.Yes | No |
| **High availability** | Yes | Yes | Yes |
| **High scalability** | Yes | Yes | Yes |
| **Single point of failure** | No | Yes | N/A |
| **Open source** | Yes | Yes | No |
| **Versioning** | Yes (External timestamp at query level, but no built-in) | Yes (built-in timestamp) | Yes (built-in timestamp) |
| **Indexing** | Basic primary and secondary | Basic (Secondary not by default) | Basic on map, supported by timestamps |
| **Data Processing nature** | Streaming and atomic batches | Batch processing | Batch processing |

## 4 Discussions

The rapid growth of structured or unstructured data from several data outlets, the growing realization of the limitations of relational database management systems, and the scarcity of appropriate data models, the user communities in research, academia and industry, are increasingly shifting their focus to NoSQL arena in the hope to address the questions raised as a result of data stress today. In 2007, Amazon introduced a distributed NoSQL system, called Dynamo, through a published paper, which was used by Amazon primary as its internal storage system. Amazon was one of first few big giants to begin to store majority of its principal business data in a storage management system that was not relational in nature (Leavitt, 2010). Ever since, the data science has made some advancement in the form of Big Data and efforts are still on to strengthen it further. The result of which is the presence of a good number of robust NoSQL data models today.

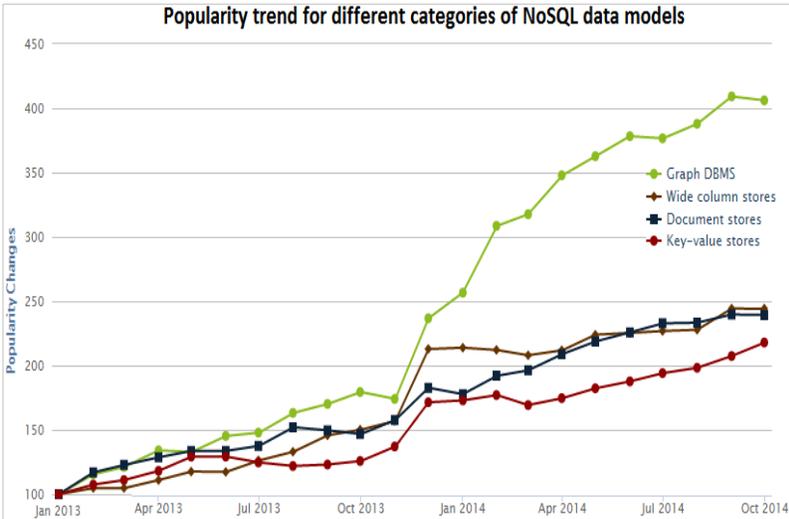

**Figure 6** Popularity trend for different categories of NoSQL data models (Source: solidIT, 2014)

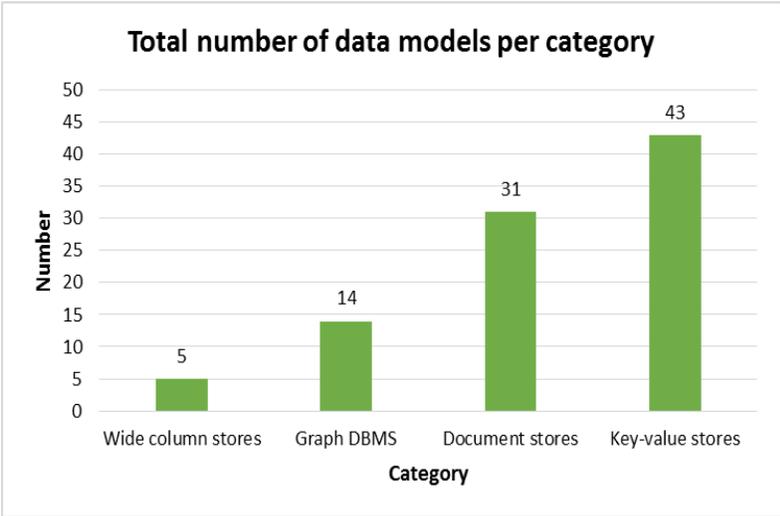

**Figure 7** Total number of data models per category

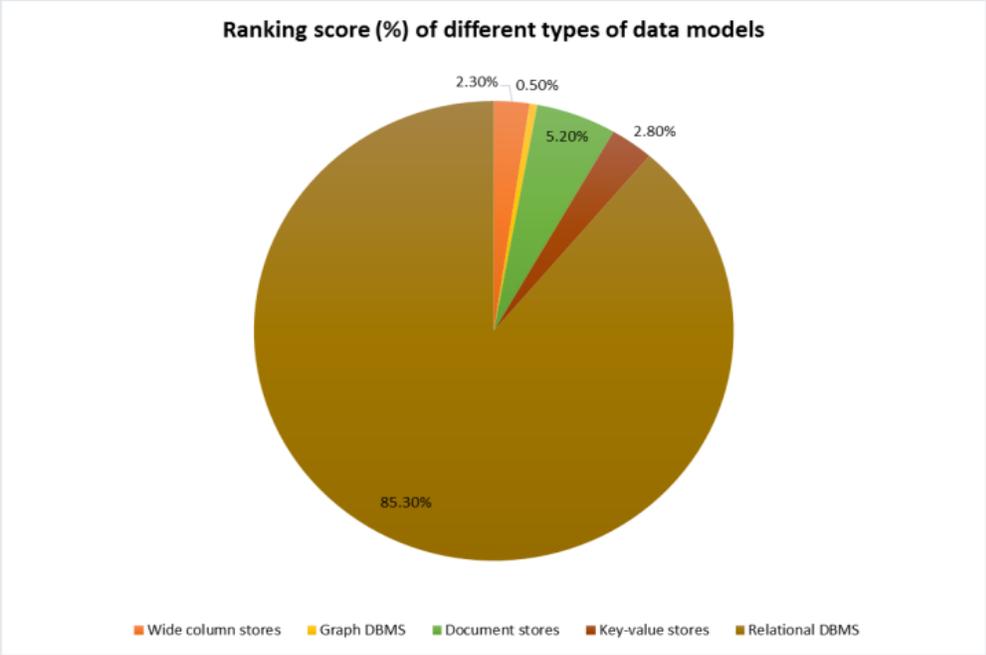

**Figure 8** Percentage ranking of various types of data models

As opposed to their peers, the graph data models highly weigh the relations and facilitate the visual representation of information, hence are considered more human-friendly. The graph data models are very useful for those scenarios where the relationships in the data have more weightage than the data itself and example includes social network representation and traversal, forensic investigations conclusion, etc.

Figure 7 demonstrates the number of data models available today for each individual category of the NoSQL, discussed in this paper. The NoSQL data models process the data at a faster rate than the relational data models, which usually are used in commercial domains that require higher precision for the transactions processes. Therefore, the relational data models rigidly comply with ACID (atomicity, consistency, isolation, durability) constraints on every piece of data and this makes the relational data models slower. Most of the leading NoSQL models are flexible enough to be used to meet the application needs and usually don't pass through ACID constraints and avoid all the technical native requirements that relational data models have, hence better performance. In contrary, for those applications that require great precision, NoSQL data models may experience overheads.

Figure 8 provides the percentage ranking (pie chart) of various categories of NoSQL family, discussed in this paper along with relational data models. The demonstration suggests, though, the NoSQL data models are gaining popularity, still they are far behind than the major relational data models. The reason is that the NoSQL data models are still alien for most of the organizations. Hence, the organizations neither feel knowledgeable nor confident enough to rightly decide about which of the NoSQL data model may better serve their need. Also, the organizations remain hesitant to completely shift from its reliable legacy data models, currently being used, to NoSQL data model; however, they are ready to try the NoSQL models in their testing environment to better understand and to gain confidence. There are several open source NoSQL data models that still need to accept the challenges of management and customer supports, similar to the relational data models commercially available.

# 5 Conclusions

The thrust of this paper is to provide a fine classification and granular comparison of a wide spectrum of the NoSQL Big Data models that apparently are being considered as the potential substitutes for the predominant relational models, which fall short in handling recent Big Data movement. The rapidly growing data proliferation by various disparate potential resources such as sensors, scientific instruments, and Internet (especially the social media), has elevated the demand to acquire the ability for high scalability, robust data processing, uniform data distribution over low-end machines (commodity servers) and high availability. This consequently compels for swift transition from the legacy design of existing data models into distribution oriented design. As, today, the data landscape in the form of Big Data has completely changed from its legacy snapshots, a thorough redesign of data management methodologies is the need of the hour. As the legacy data models, which are mainly relational in nature, are incapable of handling the today's data needs, a new era in data science is commenced that gives rise to a fast emergence of a wide range of non-relational data models, which, today, are popularly known as NoSQL and/or Big Data models. In this paper, the discussed data models are classified into four different classes (document-oriented store, graph data model, key-value store, and wide column store) and compared.

In this paper, the leading representatives of four predominant categories of NoSQL Big Data model were discussed and compared on some foremost features, which were also summarized in the tables of each classification. We hope this paper will provide at least a microscopic contribution to NoSQL Big Data literature. Though, the list of salient features of NoSQL models, discussed here, certainly has scope of extension to further benchmark the performance metrics of these models. However, the challenges of the benchmarking evaluation are reserved as future work.

# References

Abidi, D. (2011). Overview of the Oracle NoSQL Database. DBMS Musings.

Anderson, J. C., Slater N., and Lehnardt J. (2009). CouchDB: The Definitive Guide (1st ed.), *O'Reilly Media*, p. 300, ISBN 0-596-15816-5.

Angles, R. and Gutierrez, C. (2008). Survey of graph database models. *ACM Computing Surveys*, 40(1), 1.
Bell, D. (2013). UML basics: An introduction to the Unified Modeling Language. *IBM developerWorks*. [Online] http://www.ibm.com/developerworks/rational/library/769.html (Last accessed on Nov 12, 2013)

Bell, G., Hey, T., Szalay,A. (2009). Beyond the data deluge, Science 323 (5919), 1297–1298.

Berezecki, M., Frachtenberg, E., Paleczny, M., & Steele, K. (2011). Many-core key-value store. *In Proceedings of International Green Computing Conference and Workshops (IGCC),* pp. 1-8, Florida, USA.

Bertino, E., Beng, C. O., Ron, S.D., Kian, L.T., Justin, Z., Boris, S., and Daniele, A. (2012). Indexing techniques for advanced database systems. Springer Publishing Company, Incorporated.

Borthakur, D. (2007). The Hadoop Distributed File System: Architecture and Design. *Hadoop, The Apache Software Foundation*. [Online] http://hadoop.apache.org/docs/r0.18.0/hdfs_design.pdf (last access on Sept 8th, 2013)

Brown, M.C. (2011). Getting Started with CouchDB (1st ed.), *O'Reilly Media*, p. 50, ISBN 1-4493-0755-8.

Burrows, M. (2006). The Chubby lock service for loosely coupled distributed systems. *In Proceedings of the 7th symposium on Operating systems design and implementation*, pp. 335-350, USENIX Association Berkeley, CA .

Carino, F., and Sterling, W. M. (1998). Method and apparatus for extending existing database management system for new data types. U.S. Patent No. 5,794,250.

Cattell, R. (2011). Scalable SQL and NoSQL data stores. *ACM SIGMOD Record*, 39(4), 12-27.

Chang, F., Dean, J., Ghemawat, S., Hsieh, W. C., Wallach, D. A., Burrows, M., Chandra, T., Fikes, A., and Gruber, R. E. (2006). Bigtable: a distributed storage system for structured data. *In Proceedings of the 7th Conference on USENIX Symposium on Operating Systems Design and Implementation,* Volume 7 , Seattle, WA, November 06 - 08.

Chodorow, K. and Dirolf, M. (2010). MongoDB: The Definitive Guide (1st ed.), *O'Reilly Media*, p. 216, ISBN 978-1-4493-8156-1.

Collins, G., Sheehy, J., Patrick, B., Gross, A. (2010). Riak: An Open Source Scalable Data Store. Basho Technologies. [Online] http://basho.com/ (Last accessed on Nov 30, 2013)

Copper, J. (2009). How Entities and Indexes are Stored. *Google App Engine, Google Code*.

Cordes, K. (2007). *YouTube Scalability (talk), Their new solution for thumbnails is to use Google's BigTable, which provides high performance for a large number of rows, fault tolerance, caching, etc. This is a nice (and rare?) example of actual synergy in an acquisition.*

DeCandia, G. et al. (2007). Dynamo: amazon's highly available key-value store. *Proceedings of twenty-first ACM SIGOPS symposium on Operating systems principles*, Stevenson, Washington, USA.

Dimiduk, N., and Khurana, A. (2012). HBase in Action (1st ed.). *Manning Publications*. p. 350. ISBN 978-1617290527.

Eifrem, E. (2009). Neo4j-The Benefits of Graph Databases. O'reilly' OSCON-Open Source Convention. [Online] http://www.oscon.com/oscon2009/public/schedule/detail/8364 (Last accessed on Nov 17, 2013).

Fan, K. (2010). Suvey on Nosql. *Programmer, Vol (6)*: pp.76-78.

Furner, J. (2003). Little Book, Big Book: Before and After Little Science, Big Science: A Review Article, Part I. Journal of Librarianship and Information Science 35 (2): 115–125. doi:10.1177/0961000603352006. Retrieved 2014-02-09.

Garulli, L. (2012). OrientDB. Orient Technologies [Online] http://www.orientdb.org/luca-garulli.htm/ (Last accessed on Sept 27, 2013)

Ghemawat, S., Gobioff, H., and Leung, S. T. (2003). The Google File System. *19th Symposium on Operating Systems Principles (conference)*, Lake George, NY: The Association for Computing Machinery.

Google Inc. (2001). Google Earth. [Online] http://www.google.com/earth/ (Last accessed on Sept 11, 2013)

Google Inc. (2003). The Story of Blogger. Blogger.com. [Online] http://www.blogger.com/home (Last accessed on Sept 11, 2013)

Google Inc. (2004). Google Books. [Online] http://books.google.com/ (Last accessed on Oct 31, 2013)

Gremlin. (2013). Gremlin Query Language. GitHub.com (Last online accessed on September 26, 2014)

Han, J., Song, M., and Song, J. (2011). A Novel Solution of Distributed Memory NoSQL Database for Cloud Computing. *10th IEEE/ACIS International Conference on Computer and Information Science,* Sanya, China.

Hilbert, M., López, P. (2011). The world's technological capacity to store, communicate, and compute information, Science 332 (6025), 60–65.

Hoff, T. (2009). Neo4j - a Graph Database that Kicks Buttox. *High Scalability. Possibility Outpost*. [Online] http://highscalability.com/neo4j-graph-database-kicks-buttox. (Last accessed on February 17, 2010).

IBM, 2012. *What is Big Data ? Bringing Big Data to the enterprise*. [Online] http://www-01.ibm.com/software/data/bigdata/

Jin, A., Cheng, C., Ren, F., and Song, S. (2011). An index model of global subdivision in cloud computing environment. *19th International Conference on Geoinformatics*, pp. 1-5.

Johnson, R. (2010). Neo4j Cypher Refcard 1.9. [Online] http://docs.neo4j.org/refcard/1.9/ (Last accessed on Nov 12, 2013)

Lakshman, A. and Malik, P. (2011). *The Apache Cassandra Project*. [Online] http://cassandra.apache.org/.

Larry, E. (2012). Oracle NoSQL Database. An Oracle White Paper. [Online] http://www.oracle.com/technetwork/products/nosqldb/learnmore/nosql-wp-1436762.pdf (Last accessed on Nov 12, 2013)

Leavitt, N. (2010). Will NoSQL databases live up to their promise?. Computer, 43(2), 12-14.

OSGi Alliance (2010). Semantic Versioning. *Technical Whitepaper, Revision 1.0* [Online] http://www.osgi.org/wiki/uploads/Links/SemanticVersioning.pdf

Ranger, C., Raghuraman, R., Penmetsa, A., Bradski, G., and Christos, K. (2007). Evaluating MapReduce for Multi-core and Multiprocessor Systems. *Proceedings of the 2007 IEEE 13th International Symposium on High Performance Computer Architecture,* p.13-24, February 10-14, 2007 [doi>10.1109/HPCA.2007.346181].

Robert, H. (2012). It's time for a new definition of Big Data. *MIKE2.0: The open source standard for Information Management*. [Online] http://mike2.openmethodology.org/

Shaughnessy, S. T. (2012). Database system providing high performance database versioning. U.S. Patent No. 8,117,174. Washington, DC: U.S. Patent and Trademark Office.

Schadt, E. E., Linderman, M. D., Sorenson, J., Lee, L. & Nolan, G. P. (2010). Computational solutions to large-scale data management and analysis. Nature Rev. Genet. 11, 647–657.

solidIT. (2014). DB-ENGINES: *Knowledge Base of Relational and NoSQL Database Management Systems*. [Online] http://db-engines.com/en/ home (Last accessed on Sept 3, 2014)

Suryavanshi, R., Yadav, D. (2012). Modeling of Multiversion Concurrency Control System Using Event-B. *In Proceedings of the Federated Conference on Computer Science and Information Systems pp. 1397–1401,* Warsaw, Poland.

Tiwari, S. (2011). Using Oracle Berkeley DB as a NoSQL Data Store. *Oracle*. [Online] *http://www.oracle.com/technetwork/articles/cloudcomp/berkeleydb-nosql-323570.html*

Villars, R. L., Olofson, C. W. and Eastwood, M. (2011). Big Data: What it is and why you should care. *IDC White Paper*. Framingham, MA: IDC.

Watters, A. (2010). The Age of Exabytes: Tools and Approaches for Managing Big Data (Website/Slideshare). *Hewlett-Packard Development Company*.

Weiss, R., Zgorski, L. J. (2012*). Obama Administration Unveils "BIG DATA" Initiative: Announces $200 Million In New R&D Investments*. [Online] http://www.whitehouse.gov/sites/default/files/microsites/ostp/big_data_press_release.pdf

Wetherell, C. (2005).Google Reader:Two weeks.Reader is using Google's BigTable in order to create a haven for what is likely to be a massive trove of items. [Online] http://googlereader.blogspot.com/2005/10/google-reader-two-weeks.html (Last accessed on Sept 4, 2013)

Whitchcock, A. (2005). Google's BigTable. *There are currently around 100 cells for services such as Print, Search History, Maps, and Orkut*.